\documentclass[usegraphicx,usenatbib,onecolumn,doublespace]{mn2e}
\usepackage{bm}

\begin{document}

\title[Gravitational Lensing Effect on 21 cm Fluctuations]
{Gravitational Lensing Effect on the Two-Point Correlation
  Function of Hot Spots in 21 cm fluctuations}

\author[Tashiro, H. et al.]  {Hiroyuki Tashiro$^1$, Toshifumi
  Futamase$^2$ \\ $^1$ Institut d'Astrophysique Spatiale, Universit\'e
  Paris-Sud XI and CNR, Orsay, F-91405, (France)\\ $^2$ Astronomical
  Institute, Tohoku University, Sendai 980-8578, (Japan) }

\date{\today}

\maketitle

\begin{abstract}

We investigate the weak gravitational lensing effect on the two-point
correlation function of local maxima (hot spots) in the cosmic 21 cm
fluctuation map.  The intrinsic two-point function has a pronounced
depression feature around the angular scale of $\theta \sim 40$',
which depends on the observed frequency and corresponds to the scale
of the acoustic oscillation of cosmic plasma before the
recombination. It is found that the weak lensing induces a large
$w$-dependent smoothing at that scale where $w$ is the
equation of state parameter of dark energy and thus provides a useful
constraints on the dark energy property combined with the depression
angular scales on the two-point correlation function.

\end{abstract}


\maketitle

\section{Introduction}

There has been a growing interest to use the cosmic 21 cm background
fluctuations for a useful tool to study the so-called dark age of the
universe.  The emission and absorption lines of the 21 cm spin-flip
transition of neutral hydrogen produce Cosmic Microwave Background
(CMB) brightness temperature fluctuations of order $\pm 10$ mK.  By
scanning through redshift frequencies of 21 cm line, it is possible to
observe the evolution of the neutral hydrogen density with time.
Therefore, the observation of the 21 cm fluctuations is expected as a
promising probe of the reionisation history
\citep{zaldarriaga-furlanetto-hernquist-2004, mcquinn-zahn-2006}.  In
order to measure the fluctuations, there are planned low-frequency
radio arrays, for example, Mileura Wide field
Array \footnote{http://www.haystack.mit.edu/array/MWA} (MWA), Low
Frequency Array\footnote{http://www.lofar.org} (LOFAR) and Square
Kilometer Array\footnote{http://www.skatelescope.org} (SKA).
Moreover, the 21 cm cross-correlation with other observations, e.g,
CMB, large scale structures and galaxies, also studied because the
cross-correlation with complementary observation gives more
information than their respective auto-correlations
\citep{alvarez-komatsu-2006,tashiro-polarisation,adshead-furlanetto-2008,
  lidz-zahn-2008}.

The 21 cm observations also have a potential to reveal the Universe at
high redshifts before the reionisation epoch
\citep{madau-meiksin-rees-1997, tozzi-madau-2000}.  The physics of the
21 cm fluctuations at high redshifts is understood better than at
lower redshifts around the reionisation epoch (for a detailed review,
see \citealt{lewis-challinor-2007}).  The 21 cm fluctuations at high
redshifts trace the matter power spectrum and are calculated in the
linear theory.  The observations of the acoustic oscillations in the
21 cm fluctuations are expected to be a probe of the composition and
geometry of the Universe \citep{barkana-loeb-2005, mao-wo-2008}.

In addition to these primordial fluctuations, there exist secondary 21
cm fluctuations.  Modification by weak gravitational lensing is
considered as one of the main sources of secondary fluctuations
\citep{mandel-zaldarriaga-2006}.  The path of the emitted 21 cm photon
is perturbed by weak gravitational lensing of large scale structures.
Gravitational lensing modifies the primordial 21 cm fluctuations as in
the context of CMB.  However, in contrast to the CMB case where there
is only one source plane at the last scattering surface, the
redshifted 21 cm fluctuations provide excellent sources for
gravitational lensing because of the existences of many different
source planes.

In this paper, we investigate the weak lensing effect on the 21cm
fluctuations.  In order to reconstruct matter density fluctuations
along the 21 cm photon path, some authors have been studied the effect
\citep{mandel-zaldarriaga-2006,zahn-zaldarriaga-2006,
  metcalf-white-2007,lu-pen-2008}.  For example, Mandel and Zeldarriga
studied the effect on the angular power spectra as well as the
intrinsic, three-dimensional power spectra of the 21cm fluctuations
during the era of reionisation and the effect on both spectra is less
than a percent in the interesting scales
\citep{mandel-zaldarriaga-2006}.  Metcalf and White studied the lensed
shear map power spectrum in 21cm fluctuations and pointed out the
potential of producing high resolution, high signal-to-noise images of
the cosmic mass distribution \citep{metcalf-white-2007}.

Here we are interested in the two-point correlation function before
reionisation.  If adopting the Gaussian assumption for the primordial
fluctuation field, it is known that the peak statistics can provide
additional information about intrinsic distribution of hot spots that
those pairs have some characteristic separation angles
\citep{heavens-sheth-1999}.  In particular there is a pronounced
depression feature in the two-point function around the angular scale
of $\theta \sim 40$' depending on the observed frequency.  The weak
lensing then redistributes hot spots in the observed 21 cm fluctuation
maps from the intrinsic distribution.  We found that the effect has
large influence in the depression feature, typically several percent
contrary to the lensing effect on the power spectra.  Moreover, the
effect does not very much depend on the details of the nonlinear
structure formation.  Thus the detailed observation of the redshift
dependence of the weak lensing effect as well as of the angular
position in the two-point function offer a very useful mean of
investigating dark energy equation of state.

This paper is organised as follows.  In Sec.~II, we calculate and
discuss the two-point correlation function of hot spots in the 21 cm line
fluctuations.  In Sec.~III, we give the formalisms the gravitational
lensing effect on two-point correlation function.  In Sec.~IV, we
present the result of lensed two-point correlation function and the
effect of dark energy equation of state.  Finally, we conclude in
Sec.~V.  As the fiducial cosmological model in this paper, we assume
the $\Lambda$CDM cosmology with cosmological parameters, $h=0.7$
($H_0=h \times 100~{\rm km ~s}^{-1}~{\rm Mpc}^{-1}$), $\Omega_m=0.3$,
$\Omega_\Lambda=0.7$ and $\sigma_8=0.8$.

\section{two-point correlation function of hot spots in the 21 cm line fluctuations}

In the case of the CMB temperature anisotropy, two-point correlation
functions of hot spots can be calculated from the angular power
spectrum of the CMB temperature anisotropy under the Gaussian
assumption \citep{heavens-sheth-1999}.  Since it is a good assumption
that the 21 cm fluctuations before the reionisation have a Gaussian
statistic, we can calculate two-pint correlation functions of hot
spots in the 21 cm fluctuations from their angular power spectrum in
the same way as in the case of the CMB temperature anisotropy.

\subsection{Angular power spectrum of the 21 cm fluctuations}

The observed 21 cm fluctuations at $\lambda = 21(1+z_{\rm obs}) ~{\rm
  cm}$ can be written as \citep{zaldarriaga-furlanetto-hernquist-2004}
\begin{equation}
T(\hat{\bm{n}},z_{\rm obs})=T_{21}(z_{\rm obs})\int_0 ^{\eta_0} 
d\eta W_{21}(\eta_{\rm obs}-\eta)\psi_{21}
(\hat{\bm{n}},\eta),
\label{eq:21-cmflu}
\end{equation}
where $\eta$ is the conformal time, the subscripts obs and 0 mean the
values at the redshift $z_{\rm obs}$ and the present time,
respectively, and $W_{21}(\eta_{\rm obs}-\eta)$ is a response function
which characterises the bandwidth of an experiment and normalised as
$\int_{-\infty}^{\infty} d x W_{21}(x)=1$.  In this paper, we assume
that $W_{21}(\eta_{\rm obs}-\eta)$ is the delta function for
simplicity.  In Eq.~(\ref{eq:21-cmflu}), $T_{21}(z)$ is a
normalisation constant which is given by
\begin{equation}
T_{21}(z)\simeq 23~{\rm mK}~\left(\frac{\Omega_bh^2}{0.02}\right)
\left[\left(\frac{0.15}{\Omega_mh^2}\right)
\left(\frac{1+z}{10}\right)\right]^{1/2},
\end{equation}
and $\psi_{21}$ is the dimensionless brightness temperature written as
\begin{equation}
\psi_{21}(\hat{\bm{n}},\eta)\equiv x_{\rm H}(\hat{\bm{n}},\eta) 
[1+\delta_b(\hat{\bm{n}},\eta)]
\left[1-\frac{T_{\rm cmb}(\eta)}{T_{\rm s}(\hat{\bm{n}},\eta)}\right],
\label{eq:psi21}
\end{equation}
where $\delta_b$ is the baryon density contrast, $x_{\rm H}$ is the
fraction of neutral hydrogen, $T_{\rm s}$ is the spin temperature of
neutral hydrogen and $T_{\rm cmb}$ is the CMB temperature.  For
simplicity, we assume that the fraction of neutral hydrogen is $x_{\rm
  H}=1$ because the redshifts we are interested in are before the
reionisation epoch.  We also take another assumption that the gas
temperature, $T_{\rm g}$, is heated by stars or QSOs, and much higher
than $T_{\rm cmb}$.  In this case, $T_{\rm s}$ is well coupled to
$T_{\rm g}$ and $T_{\rm s} \gg T_{\rm cmb}$.  As a result,
Eq.~(\ref{eq:psi21}) does not have the dependence on $T_{\rm s}$.
Under these assumptions, the 21 cm fluctuations are determined only by
the baryon density contrast.

Note that the second assumption, $T_{\rm s} \gg T_{\rm g}$, might be
invalid at high redshifts, if there are not sufficient heat sources to
make $T_{\rm g}$ enough high.  In this case, $T_{\rm s}$ is set by the
balance between $T_{\rm g}$ and $T_{\rm cmb}$, and the absolute value
of the amplitude of the brightness temperature fluctuations depends on
$T_{\rm s}$ \citep{madau-meiksin-rees-1997}.  However, the absolute
value of the amplitude does not affect the two-point correlation
function in our analysis, since the threshold for a hot spot in this
paper is defined relative to the dispersion of the fluctuations and
the correlation function is determined by the gradient and the second
derivative of the angular power spectrum as shown in
\citet{heavens-sheth-1999}.  Therefore, our final conclusion is not
significantly changed in the case without the second assumption.

After expanding Eq.~(\ref{eq:psi21}) in Fourier series and using
Rayleigh's formula, we can obtain spherical harmonic coefficients of
the 21 cm fluctuations,
\begin{equation}
a_{\ell m}^{21}(z_{\rm obs})=4\pi (-i)^\ell \int \frac{d^3k}{(2\pi)^3}
(1+f\mu^2){{\delta}}_{b\bm{k}}
\alpha_\ell ^{21}(k,z_{\rm obs})Y_{ \ell m}^*({\bm{k}}),
\label{eq:alm21}
\end{equation}
where $Y_{ \ell m}({\bm{k}})$ is a spherical harmonic function and
$\alpha^{21}_\ell (k,z)$ is a transfer function for the 21 cm line,
\begin{equation}
\alpha^{21}_\ell (k,z)\equiv T_{21}(z) D(z)j_\ell [k(\eta_0-\eta)].
\label{eq:alpha21}
\end{equation}
Here, $j_\ell$ is the spherical Bessel function and $D(z)$ is the
linear growth factor of the density fluctuations.  We included the
factor $(1+f\mu^2)$ in order to take into account of the
redshift-space distortion by the bulk velocity fields, which is called
``Kaiser effect'', with $\mu\equiv\hat{\bm{k}}\cdot \hat{\bm{n}}$, and
$f\equiv d\ln D/d\ln a$ \citep{bharadwaj-ali-2004}.  The angular power
spectrum of the 21 cm fluctuations is obtained from
\begin{equation}
C_{21}(z_{\rm obs},\ell) = 
{\langle |a_{\ell m}^{21}(z_{\rm obs})|^2 \rangle}.
\end{equation}

\subsection{Two-point correlation function}

For calculating the two-point correlation function of hot spots in the
21 cm line fluctuations, we define the number density fluctuations of
hot spots in the 21 cm fluctuation map as
\begin{equation}
\delta n_{\rm pk}(\theta) =
\frac{n_{\rm pk}(\theta) - \bar n_{\rm pk}}{\bar n_{\rm pk}},
\end{equation}
where $n_{\rm pk}$ and $\bar n_{\rm pk}$ are the number density and
the mean number density of hot spots above a certain threshold $\nu$.
The threshold $\nu$ is written as $\nu = \Delta T_{21}/ \sigma_{21} $
with the 21 cm fluctuations $\Delta T_{21}$ and their dispersion,
\begin{equation}
\sigma_{21} \equiv \langle | \Delta T_{21} |^2 \rangle 
= \frac{1}{ (2 \pi) ^2} \int d {\bm \ell}^2  C_{21}(\ell).
\end{equation}
In what follows, we set $\nu =1$ as the threshold.

The two-point correlation function of hot spots is the ensemble
average of the number density fluctuations,
\begin{equation}
\xi_{\rm pk-pk}(\theta) \equiv \langle \delta n({\bm \theta} _1)
\delta n({\bm \theta} _2) \rangle,
\end{equation}
where $|{\bm \theta} _1 -{\bm \theta} _2| = \theta$.  The detailed
calculation of the correlation function of hot spots from the angular
spectrum is written in \citet{heavens-sheth-1999}.  Therefore, we do
not give the detailed calculation and we only show some results here.

\begin{figure}
  \begin{tabular}{cc}
   \begin{minipage}{0.5\textwidth}
  \begin{center}
    \includegraphics[keepaspectratio=true,height=50mm]{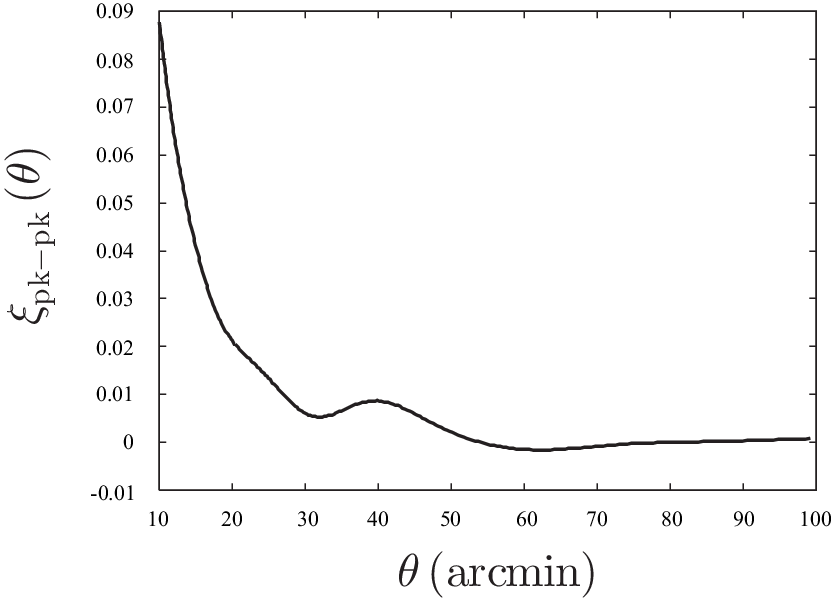}
  \end{center}
  \end{minipage}
   \begin{minipage}{0.5\textwidth}
  \begin{center}
    \includegraphics[keepaspectratio=true,height=50mm]{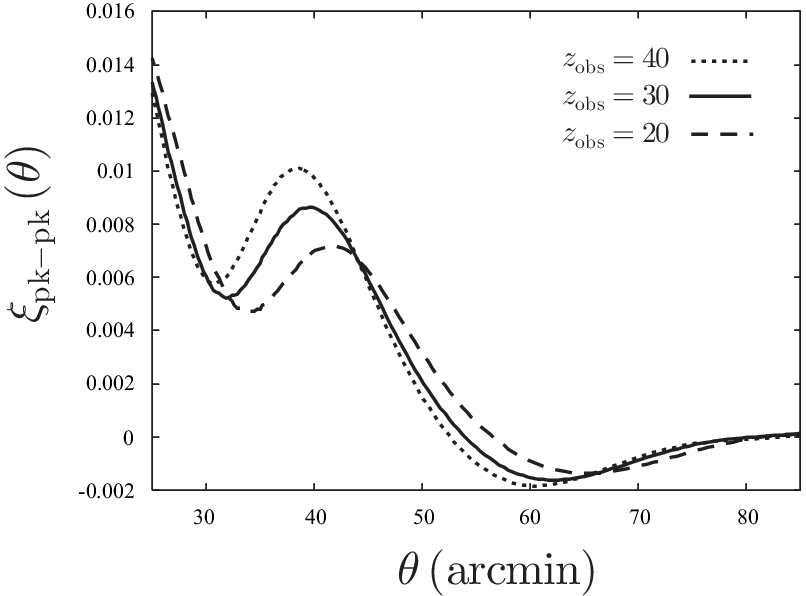}
  \end{center}
   \end{minipage}
  \end{tabular}
  \caption{ ({\it Left}) The two-point correlation function of hot
    spots in the 21 cm fluctuations.  We set $\nu=1$, and $z_{\rm
      obs}=30$.  ({\it Right}) The correlation function for different
    $z_{\rm obs}$.  The dotted, solid and dashed lines represent the
    correlation functions for $z_{\rm obs}=40$, $z_{\rm obs}=30$ and
    $z_{\rm obs}=20$, respectively.}
  \label{fig:peak21cm}
\end{figure}

The left panel in Fig.~\ref{fig:peak21cm} shows the two-point
correlation functions of hot spots in the 21 cm fluctuations.  In this
figure, we set $z_{\rm obs} =30$.  The baryonic oscillation brings the
oscillatory feature between 20 and 50 arcmin in the correlation
function.  Since the 21 cm fluctuations do not have Silk damping, the
correlation function does not damp below 10 arcmin unlike that of the
CMB temperature anisotropy.

Varying the observational wavelength, we can obtain two-point
correlation functions at different $z_{\rm obs}$.  We plot the
correlation function for different $z_{\rm obs}$ in the right panel of
Fig.~\ref{fig:peak21cm}.  When $z_{\rm obs}$ are varied from high to
low redshifts, the position of the oscillation shifts to large angle.
This is because the apparent angular diameter of the baryonic
oscillation becomes large at low redshifts.  In addition, decreasing
the amplitude of the oscillation at low redshifts is also explained by
the apparent angular diameter.  The increment of the apparent angular
diameter makes the baryonic oscillation smooth in the angular power
spectrum of the 21 cm fluctuations.  Since the two-point correlation
function are related to the gradient and second derivative of the
angular power spectrum, the smoothing of the baryonic oscillation in
low $z_{\rm obs}$ decreases the amplitude of the two-point correlation
function.

\section{gravitational lensing effect on two-point correlation functions}

During traveling to an observer, 21 cm photons are deflected by the
gravitational potential of the density fluctuations along the path
as CMB photons.  Therefore, we measure the lensed 21 cm fluctuations.  
This means that the number density of hot spots at the
direction $\theta$ on the observation map corresponds to that at
$\theta + \delta \theta $ on the source plane, where $\delta \theta$
is the deflection angle generated by the weak gravitational lensing
effect.  Therefore, the observed number density fluctuations
of hot spots, $\delta n^{\rm obs} _{\rm pk}$, are represented as
\begin{equation}
\delta n^{\rm obs} _{\rm pk} (\theta ) = \delta n (z,\theta + \delta \theta).
\end{equation}
where $\delta n (z,\theta)$ denote the number density fluctuations of
hot spots at the redshift $z$.

The dispersion of $\delta \theta$ for the angular separation $\theta$
is written as \citep{seljak-lens-1996}
\begin{eqnarray}
\sigma_{\rm GL}^2(\theta) \equiv 2^{-1}
\langle
|\delta \bm{\theta}_1-\delta \bm{\theta}_2|^2
\rangle
=\sigma_{\rm GL,0}^2(\theta)+\sigma^2_{\rm GL,2}(\theta),
\label{eq:avergl}
\end{eqnarray}
where $|\bm{\theta}_1 -\bm{\theta}_2|=\theta$
and $\langle ~\rangle$ denotes the ensemble average.
In Eq.~(\ref{eq:avergl}), $\sigma_{\rm GL,0}$ and $\sigma_{\rm
GL,2}$ are the isotropic and anisotropic contributions to the
lensing dispersion, respectively,
and given by
\begin{eqnarray}
\sigma_{\rm GL,0}^2(\theta)&=&\frac{1}{2\pi}\int \frac{d\ell}{\ell}
C_{\rm GL}(\ell)[1-J_0(\ell\theta)], \nonumber \\
\sigma_{\rm GL,2}^2(\theta)&=&\frac{1}{2\pi}\int \frac{d\ell}{\ell}
C_{\rm GL}(\ell)J_2(\ell\theta),
\label{eq:glsigma}
\end{eqnarray}
Here, $C_{\rm GL}(\ell)$ is 
the angular power spectrum of the deflection angle,
\begin{equation}
C_{\rm GL}(\ell)=9H^4_0\Omega_{\rm m}^2\int^{\eta_{\rm obs}} _{\eta_0} d \eta
a^{-2}(\eta)W^2(\eta, \eta_{\rm obs})
P_{m}\left(k=\frac{\ell}{\chi (\eta)},\eta \right).
\label{eq:lenscl}
\end{equation}
where $P_{m}$ is the matter density power spectrum and
$W(\eta,\eta_{\rm obs})$ is represented as $W(\eta, \eta_{\rm obs})=
\chi (\eta -\eta_{obs})/ \chi (\eta_{\rm obs})$ where $\chi$ is the
comoving angular diameter distance.

The effect of weak gravitational lensing on the correlation functions
of hot spots was investigated by \citet{takada-komatsu-2000} (see also
\citealt{takada-futamase-2001}).  The lensed two-point correlation
function is expressed with the dispersions of the deflection angle as
\begin{eqnarray}
\xi^{\rm obs}_{\rm pk-pk}(\theta)
&=& \langle {
\delta n^{\rm obs}(\bm{\theta}_1)
\delta n^{\rm obs}(\bm{\theta}_2) 
} \rangle_{|\bm{\theta}_1-\bm{\theta}_2|=\theta} 
=
\langle {
\delta n (\bm{\theta}_1 +\delta \bm{\theta}_1)
\delta n (\bm{\theta}_2 +\delta \bm{\theta}_2) 
} \rangle_{|\bm{\theta}_1-\bm{\theta}_2|=\theta} 
\nonumber \\
&=&
\int \frac{ d^2 {\bm \ell}}{(2 \pi)^2} 
\frac{ d^2 {\bm \ell}'}{(2 \pi)^2}
e^{i ( {\bm \ell} \cdot {\bm \theta }_1- {\bm \ell}' \cdot {\bm \theta }_2)}
\langle \delta n_ {{\bm \ell}_1} n_ {{\bm \ell}_2} \rangle
\langle e^{i ( {\bm \ell} \cdot \delta {\bm \theta }_1
- {\bm \ell'} \cdot \delta {\bm \theta }_2)} \rangle
\nonumber \\
&=& \int^\infty_0 
\frac{ \ell d \ell}{2\pi} C_{\rm pk-pk}(\ell)
\left[
\left(1-\frac{\ell ^2 }{2}\sigma^2_{\rm GL,0}(\theta)\right)J_0( \ell \theta)
+\frac{\ell ^2}{2}\sigma^2_{\rm GL,2}(\theta)J_2(l\theta)\right], 
\label{eqn:xigl}
\end{eqnarray}
where, in order to obtain the final expression,
we used the Gaussian assumption of $\delta n_{\bm \ell}$,
\begin{equation}
\langle \delta n_{\bm{\ell}}\delta n_{\bm{\ell}'} \rangle
=(2\pi)^2 C_{\rm pk-pk}(\ell) \delta^2(\bm{\ell }-\bm{\ell}'),
\end{equation}
and the following approximation,
\begin{equation}
\langle 
e^{i\bm{\ell}\cdot(\delta\bm{\theta}_1-\delta\bm{\theta}_2)}
\rangle  _{|\bm{\theta}_1 -\bm{\theta}_2|=\theta}
\simeq 1-\frac{l^2}{2} 
\left[ \sigma^2_{\rm GL,0}(\theta)
+\cos(2\varphi_l)\sigma^2_{\rm GL,2}(\theta)
\right].
\label{eq:approx-exp}
\end{equation}
The angular spectrum of the unlensed correlation function of hot spots
$C_{\rm pk-pk}$ in Eq.~(\ref{eqn:xigl}) can be related with the
unlensed correlation function $\xi^{unlensed} _{\rm pk-pk}(\theta)$ as
\begin{equation}
C_{\rm pk-pk}(\ell)=
2\pi\int_0^\pi d\theta \theta \xi^{unlensed} _{\rm pk-pk}(\theta) J_0(\ell \theta). 
\label{eq:powerxi}
\end{equation}


\section{weak lensing and dark energy}

\begin{figure}
  \begin{tabular}{cc}
   \begin{minipage}{0.5\textwidth}
  \begin{center}
    \includegraphics[keepaspectratio=true,height=50mm]{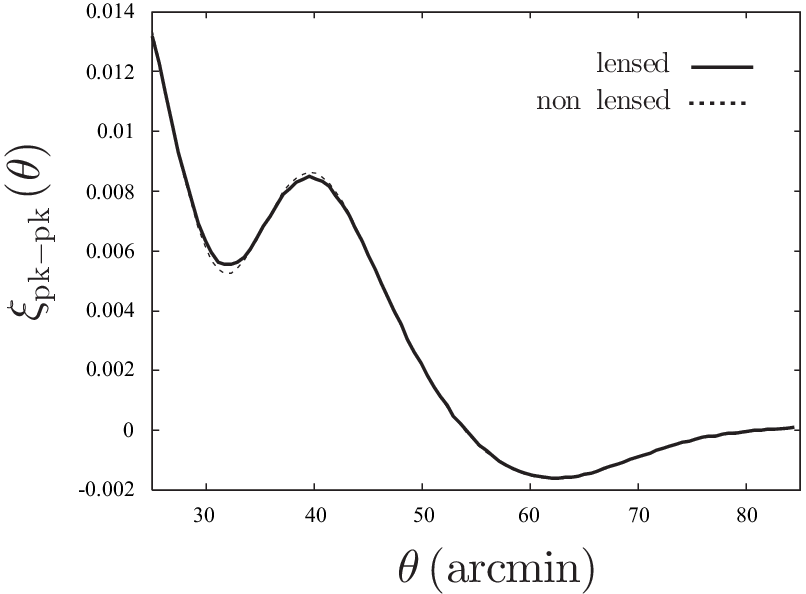}
  \end{center}
  \end{minipage}
   \begin{minipage}{0.5\textwidth}
  \begin{center}
    \includegraphics[keepaspectratio=true,height=50mm]{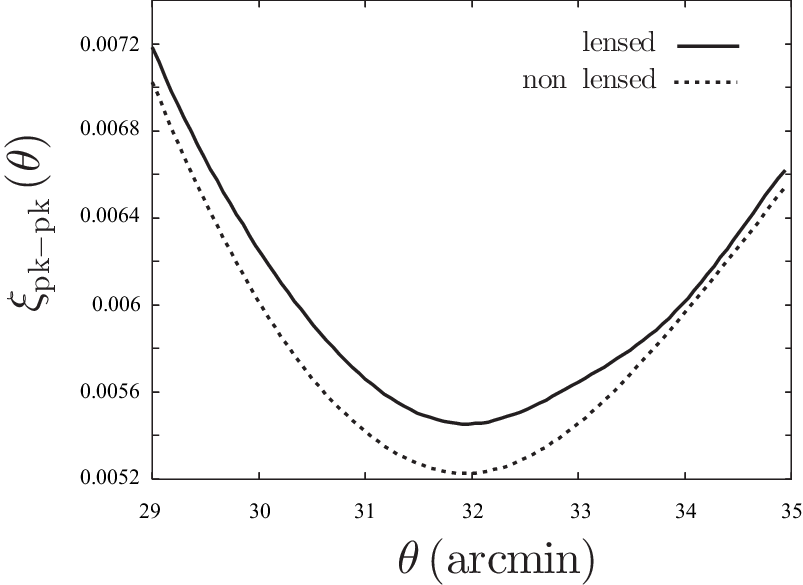}
  \end{center}
   \end{minipage}
  \end{tabular}
  \caption{The two-point correlation functions of hot spots in the 21 cm
    fluctuations. In the left panel, the solid line represents the
    lensed correlation function and the dotted line indicates the
    unlensed correlation function.  We set $z_{\rm obs} = 30$ and
    $\sigma_8=0.8$.  The right panel shows the result around 30 arcmin
    in the same case. }
  \label{fig:lensedpeak}
\end{figure}

First, we calculate the weak gravitational lensing effect on two-point
@correlation functions in the fiducial cosmological model.  The left
panel in Fig.~\ref{fig:lensedpeak} shows the lensed and the unlensed
two point correlation functions.  The effect of weak gravitational
lensing is smoothing of the oscillatory feature of two-point
correlation function as in the CMB case \citep{takada-komatsu-2000,
  takada-futamase-2001}.  Therefore, the effect of gravitational
lensing arises prominently around 30-50 arcmin where there are the top
and bottom of the oscillatory features.

In this calculation, we have not taken
into account the nonlinear evolution of matter density fluctuations,
although this gives the amplification of gravitational lensing on
small scales in the CMB case \citep{seljak-lens-1996}.  As shown in
Fig.~\ref{fig:lensedpeak}, the scale where the gravitational
lensing effect arises prominently is about more than 30 arcmin.
According to Eq.~(\ref{eq:glsigma}), most contribution of the
gravitational lensing effect at each $\theta$ comes from $C_{\rm GL}
(\ell)$ where $\ell$ corresponds to $\ell \sim \pi / \theta$.
Therefore, the gravitational lensing effect at about 30 arcmin is
sensitive to $C_{\rm GL} (\ell)$ at $\ell \sim 300$.  The discrepancy
between the linear and nonlinear density power spectra on the power
spectra of the deflection angle is made on small scales $\ell > 1000$
(see Fig.~1 in \citealt{mandel-zaldarriaga-2006}).  
Thus, neglecting the nonlinear evolution is valid in this paper.

\begin{figure}
  \begin{center}
    \includegraphics[keepaspectratio=true,height=50mm]{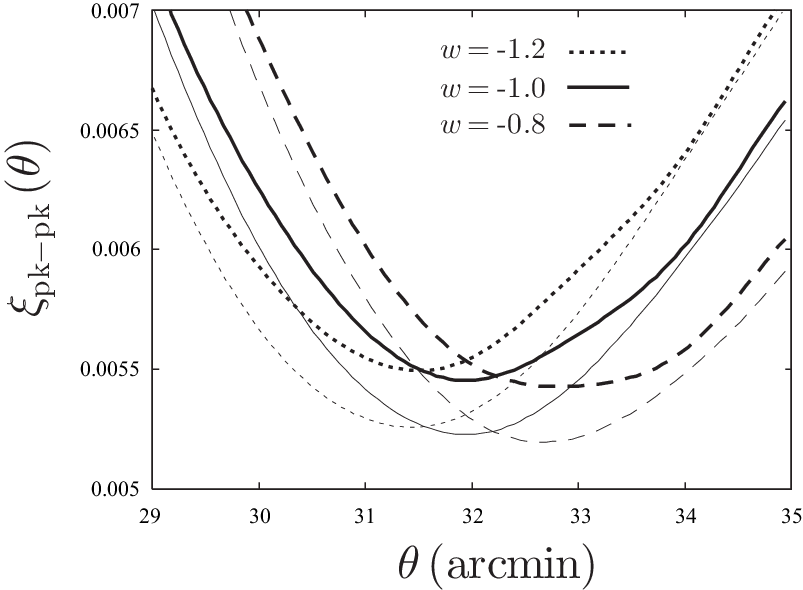}
  \end{center}
  \caption{The lensed two-point correlation functions of hot spots for
    different $w$. The dotted, solid, and dashed lines represent the
    lensed correlation functions for $w=-1.2$, $w=-1.0$, and $w=-0.8$,
    respectively.  For comparison, we also plot unlensed correlation
    function as the thin lines.  We set $z_{\rm obs} = 30$ and
    $\sigma_8= 0.8$.  }
    \label{fig:lensed-w}
\end{figure}

Next, we study the effect of the dark energy equation of state $w$ on
weak gravitational lensing in the two-point correlation function.  
Fig.~\ref{fig:lensed-w} shows the lensed two-point
correlation function for different $w$.  
One of the effects of $w$ is the shift of the position of the
oscillatory feature in the unlensed correlation function.  
Decreasing $w$ means that the acceleration of the
universe in the dark energy dominated epoch becomes high and the
distance to $z_{\rm obs}$ increases.  As a result, the
oscillatory feature in the correlation function shifts to small angle
scale as in the case of the baryonic oscillation in
CMB or galaxy redshift surveys \citep{blake-glazebrook-2003}.  
Gravitational lensing does not affect the position of the oscillatory
feature.  Therefore, even in lensed correlation functions, the
$w$-dependence of the position of the oscillatory feature is same as
in the case of unlensed correlation functions.

\begin{figure}
  \begin{tabular}{cc}
   \begin{minipage}{0.5\textwidth}
  \begin{center}
    \includegraphics[keepaspectratio=true,height=50mm]{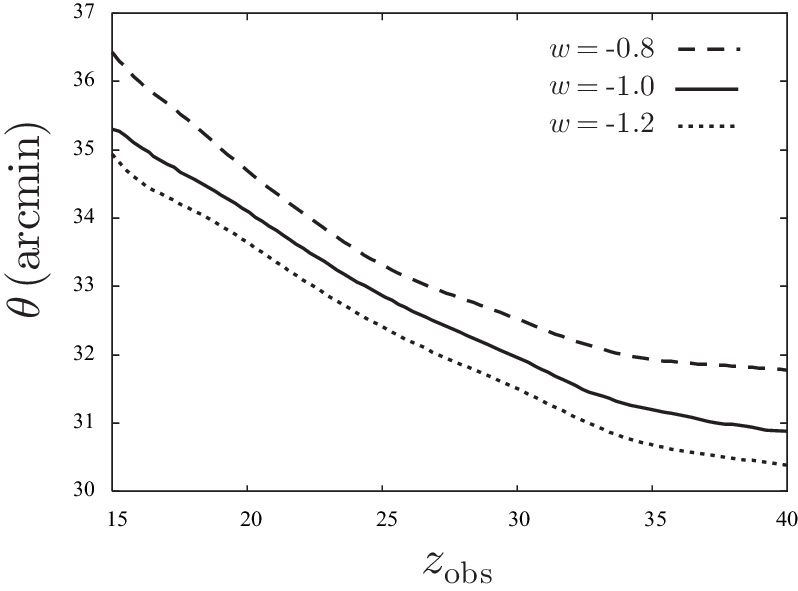}
  \end{center}
  \end{minipage}
   \begin{minipage}{0.5\textwidth}
  \begin{center}
    \includegraphics[keepaspectratio=true,height=50mm]{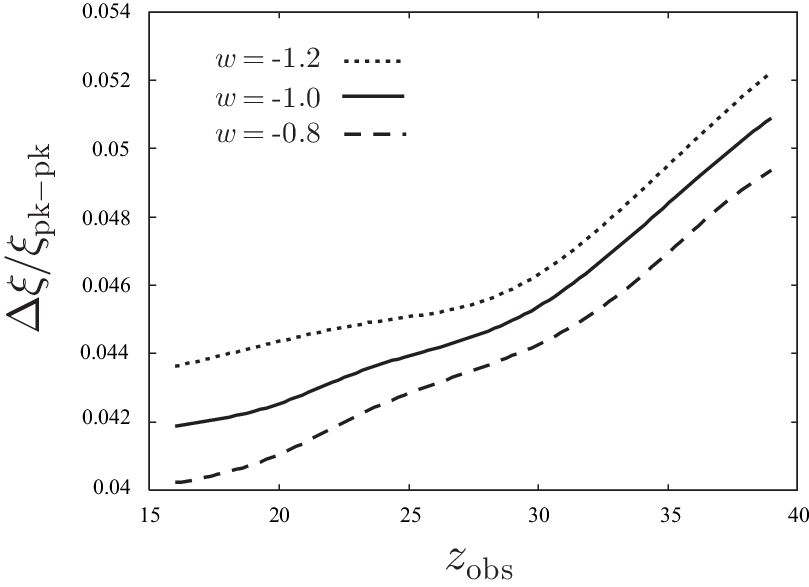}
  \end{center}
   \end{minipage}
  \end{tabular}
  \caption{ ({\it Left}) The evolution of the position of the
    oscillatory trough for different $w$.  The dotted, solid, and
    dashed lines are for $w=-1.2$, $w=-1.0$, and $w=-0.8$,
    respectively.  ({\it Right}) The evolution of fractional changes
    of the two-point correlation function for different $w$.  The
    dotted, solid, and dashed lines represent the evolutions for
    $w=-1.2$, $w=-1.0$, and $w=-0.8$, respectively.}
  \label{fig:evo-w}
\end{figure}

Varying the observational wavelength, we can obtain two-point
correlation functions at different $z_{\rm obs}$.  The left panel of
Fig.~\ref{fig:evo-w} shows the evolution of the position of the
oscillatory trough at the redshift $z_{\rm obs}$ for different $w$.  As
$z_{\rm obs}$ decreases, the apparent angle of the baryonic acoustic
oscillation becomes large.  Therefore, the position of the
oscillatory trough shifts to large scale with $z_{\rm obs}$ decreasing.

We present the evolution of the fractional change by gravitational
lensing at the trough
$\Delta \xi /\xi^{\rm unlensed} _{\rm pk-pk}$ with $ \Delta \xi = (\xi^{\rm obs} _{\rm pk-pk}
- \xi^{\rm unlensed}_{\rm pk-pk} )$, as a function
of $z_{\rm obs}$ for different $w$ in the right panel of
Fig.~\ref{fig:evo-w}. 
The deflection angle depends on the distance to the redshift $z_{\rm obs}$ 
which is the redshift of a `source
plane', as described in Eq.~(\ref{eq:lenscl}).  Decreasing $z_{\rm
  obs}$ means that the distance becomes short and the deflection angle
by gravitational lensing decreases.  Therefore, the integrated value
of Eq.~(\ref{eq:lenscl}) is smaller for low $z_{\rm obs}$ than for
high $z_{\rm obs}$.

The amplitude of $\Delta \xi
/\xi^{\rm unlensed}_{\rm pk-pk}$ depends on $w$ as shown in the right panel of
Fig.~\ref{fig:evo-w}.  Decreasing $w$ with keeping $z_{\rm
  obs}$ makes the comoving distance to $z_{\rm obs}$ large.  
As a result, integral range of Eq.~(\ref{eq:lenscl})
become large and the gravitational lensing effect is enhanced for small $w$.
The modification of the growth rate of density fluctuations by the dark
energy equation of state affects weak gravitational lensing.
However we found that this effect is
minor in our parameter region ($-1.2 < w < -0.8$), compared with the
effect of the modification of the distance to $z_{\rm obs}$.

\begin{figure}
  \begin{center}
    \includegraphics[keepaspectratio=true,height=50mm]{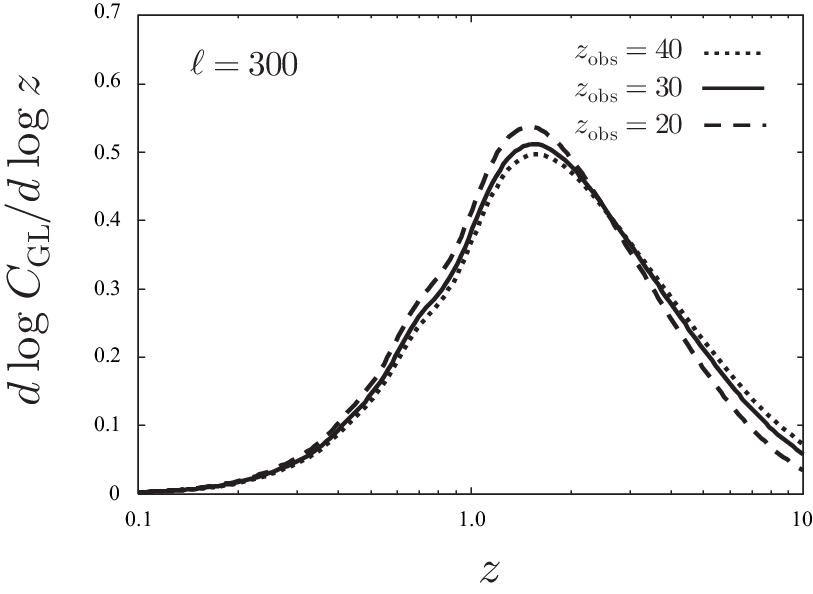}
  \end{center}
  \caption{The redshift distribution of $C_{\rm GL} (\ell)$ at $\ell=300$ 
  as a function of $z_{\rm obs}$.  
  The dotted, solid and dashed lines are
  for $z_{\rm obs}=40$, $z_{\rm obs}=30$ and $z_{\rm obs}=20$, respectively.
}
    \label{fig:clcontri-z-l300}
\end{figure}

At the last,
we investigate which redshift 
makes the most contribution to the lensing effect.
Since the gravitational lensing effect mainly arises at about 30 arcmin
in the 21 cm fluctuations between $z_{\rm obs} = 20$ and 40,
the dominant contribution to the lensing dispersion in Eq.~(\ref{eq:glsigma})
comes from $C_{\rm GL} (\ell)$ at $\ell \sim 300$.
We plot the redshift distribution of $C_{\rm GL} (\ell)$ at 
$\ell=300$ in Fig.~\ref{fig:clcontri-z-l300}.
Varying $z_{\rm obs}$ does not make the distribution change radically,
because the radial distances to the source plane at each $z_{\rm obs}$ 
are not different strongly.
For all $z_{\rm obs}$, the distribution has a peak around $z \sim 1.5$.
Therefore, gravitational lensing effect on the 21 cm fluctuations
is sensitive to the matter density fluctuations at this redshift

\section{conclusion}

We have calculated two-point correlation functions of hot spots in 21
cm fluctuations and studied the weak gravitational lensing effect on
the correlation function.  Particularly, we have examined its
possibility as a probe of the equation of state of dark energy $w$.

The two-point correlation function of hot spots in the 21 cm
fluctuations is more smoothed than that of the CMB temperature
anisotropy.  On large scales, the correlation function is very flat.
However, the amplitude of the correlation function increases toward
small scales due to the absence of Silk damping in the 21 cm
fluctuations.  The baryonic acoustic oscillation produces the
oscillatory feature around 30-50 arcmin.  By decreasing the
observation redshift $z_{\rm obs}$, the oscillatory feature shifts to
large scales.  Therefore, the position of the oscillatory feature in
multi-frequency observations plays a important role in the decision of
the cosmological parameter as the baryonic oscillation in the CMB
temperature anisotropies and large scale structures.

The weak gravitational lensing effect on the two-point correlation
function appears on the oscillatory feature.  The effect is smoothing
of the feature by smearing the baryonic oscillation without shifting
its position.  The advantage of the 21 cm fluctuation observation is
that we can obtain independent lensed maps at different redshifts.  As
$z_{\rm obs}$ decreases, the distance to the source plane becomes
small and the deflection angle by gravitational lensing decreases.  As
a result, the difference between the lensed and the unlensed
correlation functions becomes small at low $z_{\rm obs}$.

We have studied the sensitivity of the 21 cm two point correlation
function to the dark energy equation of state $w$.  The effects of $w$
on the two point correlation function appear on the shift of the
position of the oscillatory features and the smoothing by
gravitational lensing.  Since the distance to $z_{\rm obs}$ depends on
$w$, decreasing $w$ makes the position shift to small scale and the
smoothing emphasised.

The evolution of the fractional change by the gravitational lensing
effect, which is obtained from observations of different redshift
slices, is useful for the constraint on $w$.  The 21 cm fluctuations
with gravitational effect can be estimated from the linear
theory. Thus, it will be easy to compare these results with
observational data and to obtain the constraint on $w$.

We mention the comparison between the angular power spectrum and the
two-point correlation function of the 21 cm fluctuations for the
detection of the gravitational lensing effect.  According to
\citet{mandel-zaldarriaga-2006}, the fractional change by
gravitational lensing in the angular power spectrum is about $1 ~\%$.
On the contrary, the fractional change is enhanced to about $4 ~\%$ in
the correlation function.  Therefore, the two-point correlation
function is a better probe of the gravitational lensing effect than
the angular power spectrum.

Finally, we give some comments on observational aspects.  In this
paper, we focus on weak gravitational lensing on the 21 cm
fluctuations before the reionisation epoch ($z_{\rm obs} > 15$).
Therefore, we neglect any effect of reionisation process on the 21 cm
fluctuations.  Planned observational projects in the near future are
aimed at the measurement of the 21 cm fluctuations during reionisation
($z_{\rm obs} \sim 10$ for LOFAR and $z_{\rm obs} \sim 15$ for SKA).
The 21 cm fluctuations from the reionisation epoch are studied with
numerical simulations by many authors.  The angular power spectrum of
the 21 cm fluctuations depends on the reionisation process,
(e.g. \citealt{baek-dimatteo-2008}).  Therefore, in order to measure
$w$ by the weak gravitational lensing effect in the near future
observations, we must take into account the reionisation process
precisely with numerical simulations.


\begin{thebibliography}{}

\bibitem[\protect\citeauthoryear{{Adshead} \& {Furlanetto}}{{Adshead} \&
  {Furlanetto}}{2008}]{adshead-furlanetto-2008}
{Adshead} P.~J.,  {Furlanetto} S.~R.,  2008, MNRAS, 384, 291

\bibitem[\protect\citeauthoryear{{Alvarez}, {Komatsu}, {Dor{\'e}} \&
  {Shapiro}}{{Alvarez} et~al.}{2006}]{alvarez-komatsu-2006}
{Alvarez} M.~A.,  {Komatsu} E.,  {Dor{\'e}} O.,    {Shapiro} P.~R.,  2006,
  Astrophys. J., 647, 840

\bibitem[\protect\citeauthoryear{{Baek}, {Di Matteo}, {Semelin}, {Combes} \&
  {Revaz}}{{Baek} et~al.}{2008}]{baek-dimatteo-2008}
{Baek} S.,  {Di Matteo} P.,  {Semelin} B.,  {Combes} F.,    {Revaz} Y.,  2008,
  ArXiv:0808.0925

\bibitem[\protect\citeauthoryear{{Barkana} \& {Loeb}}{{Barkana} \&
  {Loeb}}{2005}]{barkana-loeb-2005}
{Barkana} R.,  {Loeb} A.,  2005, MNRAS, 363, L36

\bibitem[\protect\citeauthoryear{{Bharadwaj} \& {Ali}}{{Bharadwaj} \&
  {Ali}}{2004}]{bharadwaj-ali-2004}
{Bharadwaj} S.,  {Ali} S.~S.,  2004, MNRAS, 352, 142

\bibitem[\protect\citeauthoryear{{Blake} \& {Glazebrook}}{{Blake} \&
  {Glazebrook}}{2003}]{blake-glazebrook-2003}
{Blake} C.,  {Glazebrook} K.,  2003, Astrophys. J., 594, 665

\bibitem[\protect\citeauthoryear{{Heavens} \& {Sheth}}{{Heavens} \&
  {Sheth}}{1999}]{heavens-sheth-1999}
{Heavens} A.~F.,  {Sheth} R.~K.,  1999, MNRAS, 310, 1062

\bibitem[\protect\citeauthoryear{{Lewis} \& {Challinor}}{{Lewis} \&
  {Challinor}}{2007}]{lewis-challinor-2007}
{Lewis} A.,  {Challinor} A.,  2007, Phys. Rev. D, 76, 083005

\bibitem[\protect\citeauthoryear{{Lidz}, {Zahn}, {Furlanetto}, {McQuinn},
  {Hernquist} \& {Zaldarriaga}}{{Lidz} et~al.}{2009}]{lidz-zahn-2008}
{Lidz} A.,  {Zahn} O.,  {Furlanetto} S.~R.,  {McQuinn} M.,  {Hernquist} L.,
  {Zaldarriaga} M.,  2009, ApJ, 690, 252

\bibitem[\protect\citeauthoryear{{Lu} \& {Pen}}{{Lu} \&
  {Pen}}{2008}]{lu-pen-2008}
{Lu} T.,  {Pen} U.-L.,  2008, MNRAS, 388, 1819

\bibitem[\protect\citeauthoryear{{Madau}, {Meiksin} \& {Rees}}{{Madau}
  et~al.}{1997}]{madau-meiksin-rees-1997}
{Madau} P.,  {Meiksin} A.,    {Rees} M.~J.,  1997, Astrophys. J., 475, 429

\bibitem[\protect\citeauthoryear{{Mandel} \& {Zaldarriaga}}{{Mandel} \&
  {Zaldarriaga}}{2006}]{mandel-zaldarriaga-2006}
{Mandel} K.~S.,  {Zaldarriaga} M.,  2006, Astrophys. J., 647, 719

\bibitem[\protect\citeauthoryear{{Mao} \& {Wu}}{{Mao} \&
  {Wu}}{2008}]{mao-wo-2008}
{Mao} X.-C.,  {Wu} X.-P.,  2008, Astrophys. J., 673, L107

\bibitem[\protect\citeauthoryear{{McQuinn}, {Zahn}, {Zaldarriaga}, {Hernquist}
  \& {Furlanetto}}{{McQuinn} et~al.}{2006}]{mcquinn-zahn-2006}
{McQuinn} M.,  {Zahn} O.,  {Zaldarriaga} M.,  {Hernquist} L.,    {Furlanetto}
  S.~R.,  2006, Astrophys. J., 653, 815

\bibitem[\protect\citeauthoryear{{Metcalf} \& {White}}{{Metcalf} \&
  {White}}{2007}]{metcalf-white-2007}
{Metcalf} R.~B.,  {White} S.~D.~M.,  2007, MNRAS, 381, 447

\bibitem[\protect\citeauthoryear{{Seljak}}{{Seljak}}{1996}]{seljak-lens-1996}
{Seljak} U.,  1996, Astrophys. J., 463, 1

\bibitem[\protect\citeauthoryear{{Takada} \& {Futamase}}{{Takada} \&
  {Futamase}}{2001}]{takada-futamase-2001}
{Takada} M.,  {Futamase} T.,  2001, Astrophys. J., 546, 620

\bibitem[\protect\citeauthoryear{{Takada}, {Komatsu} \& {Futamase}}{{Takada}
  et~al.}{2000}]{takada-komatsu-2000}
{Takada} M.,  {Komatsu} E.,    {Futamase} T.,  2000, Astrophys. J., 533, L83

\bibitem[\protect\citeauthoryear{{Tashiro}, {Aghanim}, {Langer}, {Douspis} \&
  {Zaroubi}}{{Tashiro} et~al.}{2008}]{tashiro-polarisation}
{Tashiro} H.,  {Aghanim} N.,  {Langer} M.,  {Douspis} M.,    {Zaroubi} S.,
  2008, MNRAS, 389, 469

\bibitem[\protect\citeauthoryear{{Tozzi}, {Madau}, {Meiksin} \& {Rees}}{{Tozzi}
  et~al.}{2000}]{tozzi-madau-2000}
{Tozzi} P.,  {Madau} P.,  {Meiksin} A.,    {Rees} M.~J.,  2000, Astrophys. J.,
  528, 597

\bibitem[\protect\citeauthoryear{{Zahn} \& {Zaldarriaga}}{{Zahn} \&
  {Zaldarriaga}}{2006}]{zahn-zaldarriaga-2006}
{Zahn} O.,  {Zaldarriaga} M.,  2006, AstroPhys. J., 653, 922

\bibitem[\protect\citeauthoryear{{Zaldarriaga}, {Furlanetto} \&
  {Hernquist}}{{Zaldarriaga}
  et~al.}{2004}]{zaldarriaga-furlanetto-hernquist-2004}
{Zaldarriaga} M.,  {Furlanetto} S.~R.,    {Hernquist} L.,  2004, Astrophys. J.,
  608, 622

\end{thebibliography}
\end{document}